\documentclass[aps,prd, 10pt, twocolumn, nofootinbib]{revtex4-1}
\usepackage{subeqn}

\begin{document}

\title{Cosmic Censorship, Black Holes and Integer-spin Test Fields}
\author{Koray D\"{u}zta\c{s}}\email{koray.duztas@boun.edu.tr}
\author{\.{I}brahim Semiz}\email{semizibr@boun.edu.tr}
\affiliation{Bo\u{g}azi\c{c}i University, Department of Physics \\ Bebek 34342, \.Istanbul, Turkey}

\begin{abstract}
It has been argued that, starting with a slightly sub-extremal Kerr black hole instead of an extremal one, it is possible to overspin a black hole past the extremal limit and turn it into a naked singularity by sending test bodies, if one neglects radiative and self-force effects.  In this work we show that (i) an extremal Kerr black hole can not be overspun as a result of the interaction with massless integer spin test fields (scalar, electromagnetic, or gravitational), (ii) overspinning can be achieved if we start with a nearly extremal black hole instead, and (iii) for the scalar field, the argument applies to more general black holes, and also allows use of a more general field configuration. Our analysis also neglects radiative and self-force effects.
\end{abstract}

\pacs{}

\maketitle

\section{Introduction}

The deterministic nature of general relativity relies on the validity of Cosmic Censorship Conjecture which in its weak form (WCCC) states that gravitational collapse of a body always ends up in a black hole rather than a naked singularity~\cite{penrose.orig.ccc}, i.e. ``naked singularities" can not evolve starting from nonsingular initial data. Conjecturing singularities to be hidden behind event horizons without any access to distant observers enables the specification of a well defined initial value problem and thereby we can avoid the theoretical problems that could be caused by the inevitable formation of singularities in gravitational collapse.

As reasonable as it may seem, a concrete proof of the CCC has been elusive, so one way to test the validity of the conjecture is to challenge its seemingly weak spots by constructing gedanken experiments. In these experiments one envisages a black hole absorbing some particles or fields coming from infinity. The no-hair theorem \cite{mazur} in classical general relativity, which states that stationary, asymptotically flat spacetimes are uniquely described by three parameters (Mass $M$, charge $Q$, and angular momentum per unit mass $a$), guarantees that once the particles/fields are absorbed/reflected, the spacetime will settle to another black hole with new parameters $M'$, $Q'$, and $a'$. In this work we consider neutral Kerr black holes, characterized by the two parameters $M$ and $a$. The existence of the event horizon, which discriminates black holes and naked singularities, depends on an inequality involving these two parameters
\begin{equation}
M^{2} \geq a^{2}. \label {criterion}
\end{equation}
In other words, a spacetime described by the Kerr metric corresponds to a black hole, if (\ref{criterion}) is satisfied, but to a naked singularity if it is violated; in the borderline case, i.e. when  (\ref{criterion}) is saturated, the spacetime is said to describe an extremal, or a critical black hole.   We construct our thought experiments to check if we can push the initially nonsingular spacetime satisfying (\ref{criterion}) beyond the extremal limit, so that the final spacetime violates (\ref{criterion}), and describes a naked singularity. The changes in the Kerr parameters are infinitesimal, therefore one usually prefers to start the thought experiment from conditions infinitesimally close to where we would like to push the system, i.e. from the extremal black hole. The first thought experiment in this vein was constructed by Wald \cite{wald74}. He showed that particles with enough charge and/or angular momentum to overcharge/overspin a black hole either miss, or are repelled by, the black hole. This result was generalised to the case of dyonic black holes for spinless test particles \cite{hiscock} and scalar test fields \cite{dkn,toth}.

Our work is motivated by Hubeny \cite{hubeny}, and especially by Jacobson-Sotiriou (JS) \cite{Jacobson-Sot}, who started with nearly extremal black holes instead of extremal ones, and used test bodies to overcharge/overspin black holes, neglecting backreaction effects. Later Barausse et. al. considered backreaction, radiative and self-force effects for this case, and concluded that conservative self-force may prevent the particle from being captured \cite{barrauseEtal1,barrauseEtal2}. They used \mbox{MiSaTaQuWa} equations \cite{misata,quwa} for gravitational self force which are derived for a test body of mass $m$ approximated by a black hole. Similar effects were also analyzed for the case of overcharging slightly subextremal Reissner-Nordstr\"{o}m black holes~\cite{hod1,isoyamaEtal,zimmermanEtal}. There are also works extending the test-particle result to slightly subextremal Kerr-Newman black holes~\cite{saa_santarelli}, discussing quantum effects in the related contexts~\cite{hod2,matsas_daSilva,ri_saa_1,matsasEtal,ri_saa_2}, challenging WCCC with spherical shells~\cite{hod3} and claiming WCCC violation even for extremal black holes and test particles, due to higher order terms~\cite{gao_zhang}.

\section{Test field modes on the Kerr metric}

 In the present work we consider the scattering of free test fields from a Kerr black hole to see if it is possible for the black hole to turn into a naked singularity. We use massless fields of spin 0,1,2 (scalar, electromagnetic and gravitational) that satisfy Teukolsky equation for perturbations of a Kerr black hole \cite{teuk2}. Our analysis also neglects radiative and self-force effects. Restricting ourselves to neutral black holes we avoid the first order interaction of the test electromagnetic field with the electromagnetic field of the black hole, so we can treat the electromagnetic field on equal footing with scalar and gravitational fields. The thought experiment consists of a packet of waves (so that the field energy is finite) of dominant frequency $\omega$ and dominant azimuthal wave number $m$, scattering from the black hole: Initially, at $t \to -\infty$ there is no field, just the black hole. Then the field comes in from infinity, interacts with the black hole, being partially ``transmitted" and partially reflected back to infinity. The interaction changes the mass and angular momentum of the central object. As $t \to \infty$, the field decays away leaving behind (by the no-hair theorem) another Kerr spacetime, but with new values of $M$ and $a$. 

An incident wave mode can be written in Boyer-Lindquist coordinates as
\begin{equation}
\Psi (r,\theta,\phi,t)=f(\theta ,r) e^{im\phi}e^{-i\omega t}  \label{form}
\end{equation}
Due to the boundary conditions the function $f$ describes only ingoing waves at the horizon, and both ingoing and outgoing waves at infinity. In being scattered by the hole the wave causes changes $dM$ and $dJ$ in black hole parameters where $J=aM$. 

For scalar waves the stress energy tensor is 
\begin{equation}
T_{\mu \nu}=\Psi_{,\mu}\Psi_{,\nu}-\frac{1}{2}g_{\mu \nu}\Psi_{,\alpha}\Psi^{,\alpha}. \label{scalart}
\end{equation}
Since the Kerr spacetime admits two Killing vectors $\partial/\partial t$ and $\partial/\partial \phi$, the net radial flux of energy and the net radial flux of angular momentum are given by surface integrals of $-T^{r}_{\;\;t}$ and $T^{r}_{\;\;\phi}$ respectively. Therefore for scalar waves one can show that the ratio of the net flux of angular momentum to the net flux of energy across any sphere centered at the black hole is just $m/\omega$, by comparing $-T^{r}_{\;\;t}$ and $T^{r}_{\;\;\phi}$. Bekenstein states that the same ratio holds for electromagnetic and gravitational waves  by his general argument which regards the waves as a composition of many quanta with energy $\hbar \omega$ and angular momentum $\hbar m$ far from the black hole \cite{beken}:
\begin{equation}
dJ=(m/\omega)dE \label{beken}
\end{equation} 
where $dE=dM$ for the black hole. 

\section{The (sub)extremal Kerr black hole, test field modes and CCC}

We are searching for the range of frequencies that can lead to a violation of (\ref{criterion}) {\em for a given $\delta E$,} that is, for a given change in black hole mass. Let us recast the condition for WCCC violation in terms of angular momentum,
\begin{equation}
J+\delta J>(M+\delta E)^2 \label{AngmomCond}
\end{equation} 
and use the same parametrization for closeness to extremality as JS, so that we can compare with their results:
\begin{equation}
J/M^2=a/M=1-2\epsilon^2, \label{eps}
\end{equation}
where $\epsilon \ll 1$ is implied. This turns (\ref{AngmomCond}) into
\begin{equation}
\delta J  > 2\epsilon^2 M^{2} +2 M \delta E +\delta E^2. \label{jmin}
\end{equation}
Since we are dealing with waves of a given angular wave number and (dominant) frequency, we can use (\ref{beken}) to convert this to 
\begin{equation}
0  > 2\epsilon^2 M^{2} + 2 M \left(1 -\frac{\omega_{0}}{\omega} \right) \delta E +\delta E^2 \label{quad}
\end{equation}
where have defined 
\begin{equation}
\omega_{0} = \frac{m}{2M}. \label{omega_0}
\end{equation}
Conventionally, $\omega$ is taken to be positive without loss of generality. If $(\omega - \omega_{0})$ is positive, we need negative $\delta E$ to satisfy the inequality (\ref{quad}); i.e. the field must be {\em amplified} as is scatters off the black hole. This is the well-known effect of {\it superradiance} \cite{misner}, occuring for $0<\omega<\omega_{\rm sl}$. This amplification occurs for scalar, electromagnetic and gravitational waves, that is, for integer-spin fields, and for all three types, the limiting frequency is given by \cite{teuknature}
\begin{equation}
\omega_{\rm{sl}} = m\Omega = \frac{ma}{r_+^2+a^2} = \frac{ma}{2Mr_{+}}  = \omega_{0}\frac{(1-2\epsilon^{2})}{(1 + 2 \epsilon \sqrt{1-\epsilon^{2}}) } \label{super1}
\end{equation}
where $\Omega$ is the rotational frequency of the black hole, and $m$ is positive; that is, superradiance does not occur for negative $m$.

Hence we must take positive $m$ in attempting to violate CCC [for negative $m$, both $(\omega - \omega_{0})$ and $\delta E$ will be positive]. Now, to satisfy the inequality (\ref{quad}) we must have
\begin{equation}
4 M^{2} \left(1-\frac{\omega_{0}}{\omega} \right)^{2}  - 8 \epsilon^{2} M^{2} > 0. \label{discriminant}
\end{equation}

For $\omega > \omega_{0}$ we get
\begin{equation}
1-\frac{\omega_{0}}{\omega} > \sqrt{2} \; \epsilon  \Longrightarrow
\omega  >  \frac{\omega_{0}}{1-\sqrt{2}\;\epsilon} \equiv \omega_{2}
\label{firstlimits}
\end{equation}

On the other hand, for $\omega < \omega_{0}$, (\ref{discriminant}) gives
\begin{equation}
1-\frac{\omega_{0}}{\omega} < - \sqrt{2} \; \epsilon  \Longrightarrow
\omega  <  \frac{\omega_{0}}{1+\sqrt{2}\;\epsilon} \equiv \omega_{1}
 \label{secondlimits}
\end{equation}

Note that
\begin{equation}
\omega_{\rm sl} \leq  \omega_{1} \leq \omega_{0} \leq \omega_{2}, \label{omegas}
\end{equation}
where the inequalities are saturated for $\epsilon=0$, i.e. the extremal black hole. In fact, the differences smoothly go to 0 in that limit.

Now, (\ref{firstlimits}) just means $\omega  > \omega_{2}$, since $\omega_{2} \geq \omega_{0}$. But in this range, (\ref{quad}) cannot be satisfied, since both factors in the second term of (\ref{quad}) are positive.

Similarly, (\ref{secondlimits})  means $\omega  < \omega_{1}$, since $\omega_{1} \leq \omega_{0}$. Since $\omega_{\rm sl} \leq  \omega_{1}$, we can separate this range into two.
 The range $0 < \omega < \omega_{\rm sl}$ represents superradiant frequencies,  hence (\ref{quad}) cannot be satisfied, since both factors in the second term are negative. 

  However, for $\omega_{\rm sl} < \omega < \omega_{1}$, it {\em is} possible to satisfy (\ref{quad}): The first factor in the second term is negative, the second positive. For example, we may choose
\begin{eqnarray}
\bar{\omega} & = & \omega_{0}(1-\frac{3}{2}\epsilon), \nonumber \\
\bar{\delta E} & = & - M \left(1-\frac{\omega_{0}}{\bar{\omega}} \right) = \frac{3\epsilon}{2-3\epsilon}M \label{example1}
\end{eqnarray}
for $\omega$ and $\delta E$, to get
\begin{equation}
- \frac{\epsilon^{2} M^{2}}{(2-3\epsilon)^{2}} (1+12\epsilon-18\epsilon^{2})   \label{example2}
\end{equation}
for the right-hand-side of (\ref{quad}). Hence this case represents a violation of the weak CCC, given our assumptions.

We conclude that the only frequency range that can violate the WCCC is $\omega_{\rm sl} < \omega < \omega_{1}$. The fact that the interval shrinks to zero as the extreme case is approached establishes the proof that the WCCC cannot be violated by sending scalar, electromagnetic, or gravitational test fields of a certain frequency into an extremal black hole, in a limited sense generalizing~\cite{dkn,toth} to all integer-spin test fields.  

For subextremal black holes, we can parametrize the frequency in the above interval as \mbox{$\omega = \omega_{0} + (s-2) \epsilon \omega_{0}$} where $0 < s < 2-\sqrt{2}$ and we are working to first order in $\epsilon.$ Then, if $\delta E$ is chosen in the range between
\begin{equation}
\delta E_{1,2} = \left[(2-s) \mp \sqrt{(2-s)^{2}-2}\right] \epsilon M,
\end{equation}
the right-hand-side of (\ref{quad}) will be negative. The example above (\ref{example2}) corresponds to the choices $s=1/2$ and \mbox{$\delta E = (2-s)\epsilon M$}, (the central value, to first order).

It is comforting to see that $\delta E$ is of order $\epsilon M$, i.e. that $\delta E\ll M$; we also have $\delta J \ll J$; hence the test-field approximation is justified.

The only apparent problem is that at the lower end of the frequency range, that is, at $\omega = \omega_{\rm sl}$, the absorbed energy $\delta E$ passes through zero (writen as function of $\omega$ for given amplitude~\cite{teuk3}, Fig.1; after all, $\delta E$ is positive on one side and negative on the other), so that we would not be able to choose a nonzero $\delta E$ at that frequency. But the interval of the useful frequencies is an {\em open interval}, and we can always avoid $\omega_{\rm sl}$, as illustrated by the example above; and of course, one example is enough.

Therefore we have shown that there exists a combination $\omega$, $\delta J$ and $\delta E$ for any integer-spin test field incident on a slightly subextremal Kerr black hole, that can overspin the black hole into a naked singularity as long as the radiative and self force effects are neglected.

\section{Scalar waves, subextremal Kerr-Newman black holes and CCC}

For scalar fields, we can actually do better, removing the constraints of neutral black holes, massless fields and a single (dominant) frequency: In~\cite{dkn}, a cosmic censorship indicator $CCC =  M^{2} - (Q^{2} + a^{2})$ is defined for a dyonic Kerr-Newman black hole, and the change of this indicator after interaction of the black hole with a wave packet of complex massive scalar field is found to be
\begin{widetext}
\begin{equation}
\delta(CCC) =  \frac{M^{2}+a^{2}}{2\pi M} \int d\omega \sum_{l,m} f_{lm}(\omega) f^{*}_{lm}(\omega)
[\omega+\frac{eQ_{e}M-am}{M^{2}+a^{2}}]
[\omega+\frac{eQ_{e}r_{+}-am}{r_{+}^{2}+a^{2}}] B_{\omega l m} B^{*}_{\omega l m} 
\label{eq:dCCCf}
\end{equation}
\end{widetext}
where $e$ is the charge of a field quantum, $Q_{e}$ the electric charge of the black hole, $B_{\omega l m}$ the amplitude at the horizon for the transformed radial wavefunction \mbox{$U(r)=\sqrt{r^{2}+a^{2}}R(r)$} of the wavemode $\omega, l, m$ such that it has unit ingoing amplitude at infinity, and $f_{lm}(\omega)$ the coefficient showing that mode's contribution to the wave packet. The first observation is that there are no cross-terms, i.e. each mode's contribution can be calculated independently, and subsequently added. The second observation is that only frequencies between $\omega_{1}$ and $\omega_{2}$ will make negative contribution to $\delta(CCC)$, where
\begin{equation}
\omega_{1}=\frac{am-eQ_{e}r_{+}}{r_{+}^{2}+a^{2}}\;\;\; {\rm and} \;\;\;
\omega_{2}=\frac{am-eQ_{e}M}{M^{2}+a^{2}}.
\end{equation}
In terms of the same variables, it is also possible to calculate $\delta M$ and $\delta J$:  
\begin{widetext}
\begin{eqnarray}
\delta M &=&  \frac{1}{2} \int d\omega \sum_{l,m} f_{lm}(\omega) f^{*}_{lm}(\omega)
[\omega+\frac{eQ_{e}r_{+}-am}{r_{+}^{2}+a^{2}}] \, \omega \, B_{\omega l m} B^{*}_{\omega l m} \label{scalardeltaM} \\
\delta J &=&  \frac{1}{2} \int d\omega \sum_{l,m} f_{lm}(\omega) f^{*}_{lm}(\omega)
[\omega+\frac{eQ_{e}r_{+}-am}{r_{+}^{2}+a^{2}}] \, m \, B_{\omega l m} B^{*}_{\omega l m}. \label{scalardeltaJ}
\end{eqnarray}
\end{widetext}

Equation (\ref{scalardeltaM}) allows us to identify $\omega_{1}$ as the superradiance limit; consistent with the paragraph after (\ref{omega_0}). Of course, the main result of~\cite{dkn} is that for extremal black holes, $\omega_{1}$ and $\omega_{2}$ coincide, hence $CCC$ cannot be decreased from its value of zero. But now, let us consider a slightly subextremal black hole. Such a black hole will be characterized by a small positive initial value of $CCC$, say $\varepsilon^{2} M^{2}$ in a parametrization similar to that of JS. Then, to turn this black hole into a naked singularity, we need to have $\delta(CCC) < - \varepsilon^{2} M^{2}$. This could be accomplished by a wave packet with frequencies dominant between $\omega_{1}$ and $\omega_{2}$, say a Gaussian distribution with peak near the average of $\omega_{1}$ and $\omega_{2}$ {\em for each $m$} (tails may extend outside the interval, as long as the interval is dominant). Then we can estimate for typical values of $f_{lm}(\omega)$:
\begin{equation}
|f_{lm}|^{2} \gtrsim \frac{8 \pi \varepsilon^{2} M^{3}}{(M^{2}+a^{2}) \Delta\omega \, (\omega_{2}-\omega_{1})^{2} |B|^{2}} \alpha_{lm}
\end{equation}
where $\Delta\omega$ is the width of the frequency distribution, $|B|$ the typical magnitude for $B_{\omega l m}$, and $\alpha_{lm}$ a factor of order unity or smaller, describing the contribution of modes labeled by $l$ and $m$ to $\delta(CCC)$. But, $(\omega_{2}-\omega_{1})$ must depend on $\varepsilon$, considering that it vanishes in the extremal case. We find to first order in $\varepsilon$ 
\begin{equation}
(\omega_{2}-\omega_{1}) \approx \frac{\varepsilon M}{(M^{2}+a^{2})^{2}} [2 M a m - e Q_{e} (M^{2}+a^{2}) ]. 
\end{equation}
Since $\Delta\omega$ must also be similar to $(\omega_{2}-\omega_{1})$, we have $|f_{lm}|^{2} \sim 1/\varepsilon$. This seems surprising at first, but one should keep in mind that the single-mode case also corresponds to $|f_{lm}|^{2} \sim \delta(\omega-\omega_{0})$. 

These results give us estimates for contributions of same modes to $\delta M$ and $\delta J$, again to first order in $\varepsilon$ :  
\begin{eqnarray}
\delta M_{lm} & \gtrsim & \frac{\varepsilon\pi M (am-eQ_{e}M)}{am-\frac{eQ_{e}}{2M}(M^{2}-a^{2}) } \alpha_{lm}  \\
\delta J_{lm} & \gtrsim & \frac{\varepsilon \pi  M (M^{2}+a^{2}) m}{am-\frac{eQ_{e}}{2M}(M^{2}-a^{2}) } \alpha_{lm}  
\end{eqnarray}
from which we can see that $\delta M \ll M $ and $\delta J \ll J$, confirming the validity of the test field approximation. Hence we have generalized the Hubeny -- JS idea~\cite{hubeny,Jacobson-Sot} to general classical black holes (not just Reissner-Nordstr\"{o}m or Kerr, but with four parameters, including magnetic charge), using tailored wave packets of general (massive, complex) scalar test fields (but no photons). We have neither assumed single-mode perturbations, nor the Bekenstein relation (\ref{beken}) initially, but were forced to use a narrow range of frequencies for each $m$ used [the absence of cross-terms in the integral expression (\ref{eq:dCCCf}) allows us to use a single positive $m$, if we want, approximately realizing the Bekenstein case].

\section{Conclusions}

We have shown that the idea of starting from slightly subextremal black holes to violate WCCC seems to work also using integer-spin test fields instead of test particles; again, if radiative and self-force effects ate neglected. Assuming Kerr black hole, and massless single-frequency-dominant fields, we can present a unified treatment for all three types of fields. In addition, for scalar fields, we can start from very general assumptions about the black hole and the field; but the requirement of WCCC violation forces the field to (a) narrow frequency range(s), after all.

\end{document}